\documentclass[11pt]{elsarticle}

\pdfoutput=1

\makeatletter
\def\ps@pprintTitle{%
  \let\@oddhead\@empty
  \let\@evenhead\@empty
  \let\@oddfoot\@empty
  \let\@evenfoot\@oddfoot
}
\makeatother

\usepackage{url}
\usepackage{breakurl}
\usepackage[breaklinks,
            colorlinks = true,
            linkcolor = blue,
            urlcolor  = blue,
            citecolor = blue,
            anchorcolor = blue]{hyperref}

\usepackage{lineno}

\usepackage{graphicx}
\usepackage[margin=1.25in]{geometry}
\usepackage[usenames,dvipsnames]{color}

\newcommand\acp{\begin{center}
\rule[-0.2in]{\hsize}{0.01in}\\\rule{\hsize}{0.01in}\\
\vskip 0.1in Submitted to the  Proceedings\\ 
of the African Conference on Fundamental and Applied Physics
    \vskip 0.05in
    {\it Second Edition, ACP2021, March 7--11, 2022 --- Virtual Event}\\
\rule{\hsize}{0.01in}\\\rule[+0.2in]{\hsize}{0.01in} \\
\end{center}}
\usepackage[firstpage=true]{background}
\backgroundsetup{contents={\parbox{6.5in}{\acp}}, scale=1,placement=top,opacity=1,color=black,position={3.25in,1.2in}}
\usepackage{fancyhdr}
\fancypagestyle{plain}{%
  \fancyhf{}%
  \fancyhead[C]{}
  \fancyfoot[C]{\thepage}
}

\fancypagestyle{empty}{%
  \fancyhf{}%
  \fancyhead[C]{{\it ACP2021, March 7--11, 2022 --- Virtual Edition}}
  \fancyfoot[C]{\thepage}
}
\begin{document}

\begin{frontmatter}


\title{X-ray technological irradiation for TID studies on silicon sensor and electronic devices in a medical facility}
\author[add1]{Benedetto Di Ruzza\corref{cor1}}
\ead{benedetto.diruzza@gmail.com}
\cortext[cor1]{Corresponding Author}
\address[add1]{TIFPA-INFN, Trento Institute for  Fundamental Physics and Applications\\(Italian National Institute for Nuclear Physics),\\Via Sommarive 14, 38123, Trento, (Italy)\\
ORCID iD: \href{https://orcid.org/0000-0001-9925-5254}{0000-0001-9925-5254}
}
%
%
%
%
\begin{abstract}
\noindent
Technological tests of Total Ionizing Dose effects are required not only for silicon particle sensors developed in high energy physics experiments, but also for electronic devices and semiconductor elements used in commercial, automotive and space applications. Using x-ray irradiators, these technological tests and studies can be performed not only in facilities explicitly built for this mission, but also in medical or biological research facilities when some minima requirements are satisfied. Generally this irradiation can be performed without interfering with the medical and biological activities of the facility.
In this article will be described the minimum instrumentation required for this type of studies and will be given a detailed description of the preparation and realization of a SiPM x-ray irradiation campaign for TID characterization realized in the Italian TIFPA-INFN Trento Center Laboratory in May 2021, using instruments originally realized for medical and biological irradiation.
\end{abstract}
\end{frontmatter}
%
%
%
%
\section{Introduction}
\label{sec:intro}
\noindent
The effects of Total Ionizing Dose (TID)~\cite{CERN1} on silicon sensors or devices built with semiconductor junctions (processors, RAMs, controllers, FPGA) is the modification of the electronic properties of these elements due to the exposition of these elements to a large amount of high energy electromagnetic waves (mainly x-ray). This exposition can not only change the fundamental electronic properties of these elements but also compromise completely functionality of the element itself. 
Due to the relevance that silicon sensors, microprocessors, RAMs, FPGA and controllers have not only in scientific research (high energy physics, space research) but also in our every-day life (medical devices, GRID control systems, automotive controllers, industrial safety control systems) it is important to evaluate the maximum amount of x-ray radiation dose that can be absorbed by a system based on semiconductor elements, without changing the basic properties of the system itself.
The limit of the maximum amount of absorbed x-ray compatible with the correct functioning of an electronic system can be measured performing an x-ray irradiation of the system.
Since x-ray irradiators are often present in medical and biological research facilities and laboratories for cell samples experiments, some of these irradiators can be converted in irradiators for technological irradiation on silicon devices for TID studies without compromising of the original medical/biological purpose of the device.
\section{The TIFPA-INFN experience} \label{sec:grate}
\begin{figure}[!htbp]
\begin{center}
\includegraphics[width=\textwidth]{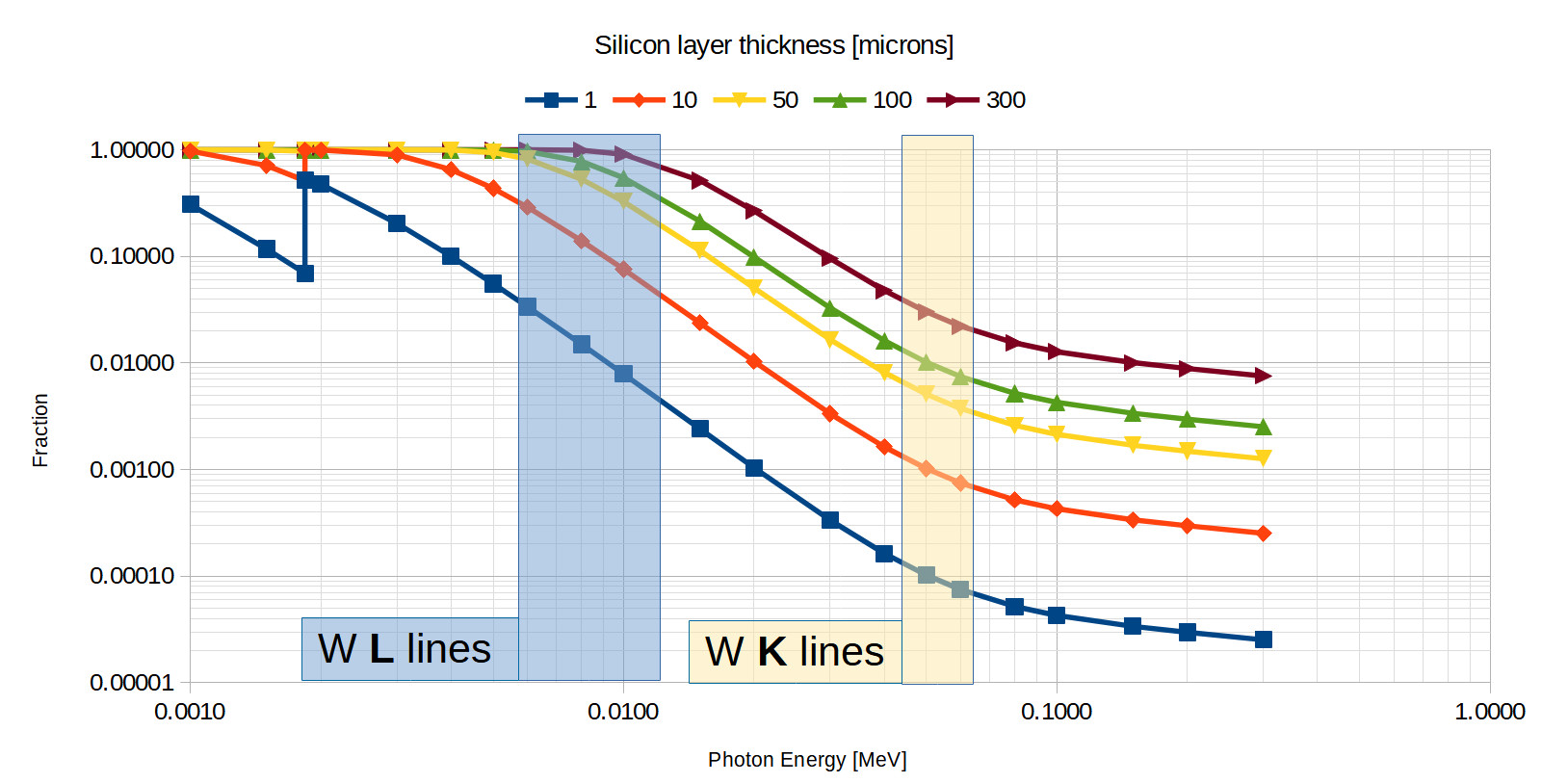}
\end{center}
\caption{Fraction of energy absorbed by a thin silicon layer for [0.001-0.300]MeV x-ray photons.}
\label{fig:pinkEle}
\end{figure}

\noindent
The Trento Institute for Fundamental Physics and Applications (TIFPA)~\cite{TIFPA1} is a research center~\cite{BDR3}\cite{BDR4} of the Italian National Institute for Nuclear Physics (INFN)~\cite{INFN1} located in Trento (Italy). This center was equipped since 2017 with an x-ray cabinet irradiator, used by TIFPA, CIBIO~\cite{CIBIO1} and other researchers of the Trento University~\cite{UNITN1}  for biological studies on cell samples.
This cabinet irradiator is based on a 3KW x-ray emission tube containing a tungsten anode. Biological irradiation are mainly performed using the K tungsten emission lines. In the TIFPA irradiator cabinet it is obtained setting the anode voltage at 195kV and using a 3mm aluminum filter to cut the low energy part of the emission spectrum.
The delivered dose is measured using a calibrated PTW ionization chamber~\cite{PTW1}.\\
The described TIFPA irradiator cannot be used as is for technological irradiation on silicon devices because the x-ray emission spectrum (tungsten K lines) is not the correct one for TID studies.
The purpose of an x-ray irradiation for TID studies is to deliver the maximum amount of energy inside the superficial thin layer of the silicon devices. In the plot in figure[\ref{fig:pinkEle}] (extracted from NIST~\cite{NIST1} data values of the mass attenuation coefficient of x-ray photons in silicon) can be observed that, in case of tungsten anode, the most efficient lines of the emission spectrum for TID studies are the L tungsten emission lines because the fraction of x-ray absorbed photons in a thin silicon layer is higher for L lines than for K lines.
The TIFPA irradiator cabinet spectrum optimization for technological irradiation was performed using the SpekPy software toolkit~\cite{SOFT1}\cite{SOFT2}. The configuration suggested by the simulation, in order to maximize the tungsten x-ray emission intensity of the L lines, was to set a 40kV anode tension and use a 0.180 mm aluminum filter.\\
It is evident at this point that an x-ray irradiator cabinet, designed for medical and biological irradiation, can be used also for technological irradiation only if the emission spectrum can be maximized in the [5-10] keV energy range.\\
In TID studies, the standard delivered dose measurement system is based on calibrated diode~\cite{CERN2}. After a comparison of the read-out of the TIFPA dose measurement system (based on a PTW ionizing chamber) vs the read-out of a dose measurement system based on calibrated diode~\cite{BDR1}\cite{BDR5}, a irradiation was performed on SiPMs prototypes designed and realized by  FBK~\cite{FBK1}. This irradiation was the last step of a SiPM radiation hardness characterization campaign described in~\cite{BDR2}.
\section{Conclusions}
\label{sec:conc}
Performing small changes in the x-ray irradiator configuration is possible to use an irradiator system, originally optimized for biological irradiation, also for technological irradiation and TID characterization of silicon sensors and electronic devices: it is required to maximize the x-ray tube emission spectrum around [5-10]keV. For a tungsten anode tube this maximization can be be realized setting the anode tension voltage at 40kV and using a 0.180 mm aluminum filter. A dose measurement system using ionization chamber can be used after the comparison of the ionization chamber read-out vs a calibrated diode dose read-out system.
\section*{Acknowledgments}
The author would like to thank Fabio Acerbi and Anna Rita Altamura (FBK) for the proficient discussions, the useful suggestions and the collaborative support in the realization of this project.


\bibliographystyle{elsarticle-num}
\bibliography{myreferences} 

\end{document}